\documentstyle[prl,preprint,aps,epsf]{revtex}
\input psfig.sty
\input{epsf}
\newcommand{\sizeplot}[6]{
	\begin{figure}[#5]
	\leavevmode \centering
	\psfig{file=#1,width=#3,height=#4,silent=}
	\caption{\label{#2} #6}
	\end{figure}}
\def\bege{\begin{equation}}
\def\ende{\end{equation}}
\def\ra{\rangle}
\def\la{\langle}

\def\bdra{\buildrel \rightarrow \over}
\draft
\begin{document}
\tightenlines
\title{\bf Quantum Field Theory of Meson Mixing }
\author{  Michael Binger and Chueng-Ryong Ji}
\address{ Department of Physics, North Carolina State University \\
          Raleigh, North Carolina 27695-8202 USA \\
          }
\maketitle
\begin{abstract}

   We have developed a quantum field theoretic framework for scalar and pseudoscalar meson
mixing and oscillations in time. The unitary inequivalence of the Fock
space of base (unmixed) eigenstates and the physical mixed eigenstates is proven and shown to lead to
a rich condensate structure. This is exploited to develop
formulas for two flavor boson oscillations in systems of arbitrary boson
occupation number. The mixing and oscillation can be understood in terms 
of vacuum condensate which interacts with the bare particles to induce non-trivial effects. 
We apply these formulas to analyze the mixing of $\eta$ with
${\eta}'$ and comment on the $K_L$ $K_S$ system. In addition, we consider the
mixing of boson coherent states, which may have future applications in the
construction of meson lasers.

\end{abstract}
%\pacs{11.15.-q,11.30.Rd,12.20.-m}
{\newpage}

{\center{\bf I. Introduction}}

The study of mixing transformations plays an important part in particle physics
phenomenology.{\cite{1}} The Standard Model incorporates the mixing of fermion fields through
the Kobayashi-Maskawa{\cite{2}} mixing of 3 quark flavors, a generalization of the 
original Cabibbo{\cite{3}} mixing matrix between the $d$ and $s$ quarks. In addition, neutrino mixing 
and oscillations are the likely resolution of the famous solar neutrino puzzle{\cite{4}}. 
In the boson sector, the mixing of $K^0$ with $\overline{K^0}$ via weak currents provided
the first evidence of $CP$ violation{\cite{5}}. The ${\eta}$ ${\eta}'$ mixing in the $SU(3)$
flavor group provides a unique opportunity for testing $QCD$ and the constituent quark model.
Furthermore, the particle mixing relations for both the fermion and boson case are 
beleived to be related to the condensate structure of the vacuum. The non-trivial nature
of the vacuum is expected to hold the answer to many of the most salient questions 
regarding confinement and the symmetry breaking mechanism.

 The importance of the fermion mixing transformations has recently prompted 
a fundamental examination of them from a quantum field theoretic perspective{\cite{6,7}}.
To our knowledge, a similar analysis in the bosonic sector has not yet been undertaken.
Moreover, the statistics of bosons and fermions are intrinsically different. Thus, the results
for boson mixing are expected to be quite different from the previous analysis of fermions. 
 That is the motivation for the present work.
 
 We begin in Section II with an investigation of the vacuum structure and the related condensation,
 using the relation between the base eigenstate and the 
 physical mixed eigenstate fields as our starting point.
 The unitary inequivalence of the associated Fock spaces is proven and an explicit formula 
 for the condensation density is derived. In Section III, the ladder operators are contructed 
 in the mixed basis. These are used to derive time dependent oscillation formulas for 
 $1$ boson states, $n$ boson states, and boson coherent states. We also show how the 
 ladder operators can be generated from two sucessive similarity transformations. Section
 IV is devoted studying specific cases in our formalism, such as the ${\eta}$ ${\eta}'$
  system. Finally, in Section V we offer some 
  concluding remarks and explore future possibilities.

{\center{\bf II. The Vacuum Structure and Condensation}}

We take the mixing of two arbitrary flavors of bosons to be given by 
\begin{eqnarray}
\Phi_{\alpha}(x) &=& \cos{\theta}{\phi}_1(x)+\sin{\theta}{\phi}_2(x)
\nonumber\\
\Phi_{\beta}(x) &=& -\sin{\theta}{\phi}_1(x)+\cos{\theta}{\phi}_2(x)
\end{eqnarray}
 where $\phi_i,i=1,2$ are solutions to the real Klein Gordon equation and
are given by 
\begin{equation} 
\phi_i(x)={\sum}_{\bdra k}
\frac{1}{\sqrt{2VE_{k,i}}}[a_{k,i}e^{-ik{\cdot}x}+a^{\dagger}_{k,i}e^{ik{\cdot}x}]. 
\end{equation}
The commutation relations are 
\begin{equation}
[a_{k,i},a^{\dagger}_{k',j}]=\delta_{kk'}\delta_{ij}
\end{equation}
from which it follows 
\bege
[\phi_i({\bdra x}),\dot{\phi}_j({\bdra y})]=i\delta({\bdra x}-{\bdra y})\delta_{ij}.
\ende
For calculational simplicity in the following we shall redefine 
$a_{k,i}{\rightarrow}a_{k,i}e^{-iE_{k,i}t}$. It is not difficult to see that
the algebra of the annihilation and creation operators remains intact and
that this redefinition does not effect any of the results we obtain.

In order to analyze the condensation density and structure of the vacuum,
we must first determine the relationship between the Fock space of
base eigenstates and the Fock space of physical mixed states. To this end
we need the unitary generator that rotates base eigenstates into physical
eigenstates :
\begin{eqnarray}
\Phi_\alpha(x) &=& G^{-1}(\theta)\phi_1(x)G(\theta)
\nonumber\\
\Phi_\beta(x) &=& G^{-1}(\theta)\phi_2(x)G(\theta)
\end{eqnarray}
 Using the Baker-Hausdorf lemma one can easily verify that 
\begin{equation}
G(\theta)=exp{\Bigl{[}}{-i\theta{\int}d^3x(\dot{\phi}_1(x)\phi_2(x)-\dot{\phi}_2(x)\phi_1(x))}{\Bigl{]}}
\end{equation}
is the generator. The commutation relations allow us to rewrite this as
$G(\theta)=e^{iS\theta}$ where
\begin{equation}
S={\sum}_{\bdra k}\frac{i}{2}[{\gamma}_-a_{-1}a_2+{\gamma}_+a_{-1}a^{\dagger}_{-2}
-{\gamma}_+a^{\dagger}_1a_2-{\gamma}_-a^{\dagger}_1a^{\dagger}_{-2}].
\end{equation}
and $\gamma_{\pm}=\sqrt{\frac{E_1}{E_2}} {\pm} \sqrt{\frac{E_2}{E_1}}$.
Here we have suppressed all of the ${\bdra {k}}$ subscripts on the ladder operators for 
notational simplicity and $a_{-1}$ stands for $a_{-k,1}$, for example. Similarly we will use 
${a_{\alpha}}$ for ${a_{k,{\alpha}}}$.

 We note that
 \begin{equation}
  _{1,2}{\la}a|{\phi}_1(x)|b{\ra}_{1,2} = $ $_{1,2}{\la}a|G(\theta)\Phi_\alpha(x)G^{-1}(\theta)
  |b{\ra}_{1,2}
  \end{equation}
  implies $G^{-1}(\theta)|b{\ra}_{1,2} {\in} \cal{H}_{\alpha,\beta}$ and 
  $|0{\ra}_{\alpha,\beta}=G^{-1}(\theta)|0{\ra}_{1,2}$. Here $|$ $ {\ra}_{1,2} {\in} {\cal{H}}_{1,2}$ and
    $|$ $ {\ra}_{{\alpha},{\beta}} {\in} {\cal{H}}_{{\alpha},{\beta}}$, where ${\cal{H}}_{1,2}$ and 
    ${\cal{H}}_{{\alpha},{\beta}}$ are the Fock space of base (unmixed) eigenstates and the Fock 
    space of physical mixed eigenstates, respectively.
  In this form we see that
  \begin{equation} 
   _{\alpha,\beta}{\la}0|0{\ra}_{1,2} = 0 
  \end{equation} 
  trivially follows. This proves the unitary inequivalence of the Fock space of base and
  physical mixed eigenstates even in the finite volume regime. For fermions, Blasone and Vitiello{\cite{6}} 
  have found that the respective Fock spaces are unitarily inequivalent only in the 
  infinite volume limit. This contrast arises because fermions have a finite number
  of states in a finite volume whereas bosons have an uncountable infinity of states in a finite volume.
   Thus, to obtain the aggregate particle behavior which manifests itself in the vacuum
   states, it is necessary to go to an infinite volume for fermions but not for bosons. 
  
  We define the number operator in the natural way, $N_{k,i}{\equiv}N_i=a^{\dagger}_ia_i$.
  The condensation density of the physical vacuum is defined as $_{\alpha,\beta}{\la}0|N_1|0{\ra}
  _{\alpha,\beta}$.
  It follows that 
 \begin{equation}
  _{\alpha,\beta}{\la}0|N_1|0{\ra}_{\alpha,\beta} =
  _{1,2}{\la}0| e^{iS{\theta}} a^{\dagger}_1 e^{-iS{\theta}} e^{iS{\theta}} a_1 e^{-iS{\theta}} |0{\ra}_{1,2},
 \end{equation} 
 where 
 \begin{equation}
 e^{iS{\theta}} a_1 e^{-iS{\theta}} = 
          a_1\cos{\theta} - \frac{\sin{\theta}}{2}({\gamma}_+a_2 + {\gamma}_-a^{\dagger}_{-2}). 
 \end{equation}
 From these we easily obtain 
 \begin{equation}
  _{\alpha,\beta} {\la}0|N_1|0{\ra} _{\alpha,\beta} =$ $ _{\alpha,\beta} {\la}0|N_{2}|0{\ra}
  _{\alpha,\beta}
  = \frac{\gamma_-^2}{4} \sin^2\theta
 \end{equation}
 Therefore, an admixture of base-eigenstate particles is found in the physical vacuum state. As we 
 will see, this condensation density becomes manifest in the boson mixing relations to be derived later.
 Note that the converse is also true. The base vacuum state contains an admixture of 
 physical eigenstate particles and the condensation density, given by
 $_{1,2}{\la}0|N_{\alpha}|0{\ra}_{1,2} =$ $_{1,2}{\la}0|N_{\beta}|0{\ra}_{1,2}$,
 is the same as above(eq.(12)). 

 {\center{\bf III.Ladder Operators and Mixing Relations}}

 The ladder operators in the mixed basis are given from eq.(5) as 
 \bege
 a_{\alpha}=G^{-1}(\theta)a_1G(\theta),
 \ende
 assuming equal masses in the two eigenstate representations, or after a simple redefinition of the operators.
 This leads to the following operators : 
 \begin{eqnarray}
 a_{\alpha} &=& a_1\cos{\theta}+\frac{\sin{\theta}}{2}({\gamma}_+a_2+{\gamma}_-a^{\dagger}_{-2})\nonumber\\
 a_{\beta}  &=& a_2\cos{\theta}-\frac{\sin{\theta}}{2}({\gamma}_+a_1+{\gamma}_-a^{\dagger}_{-1}).
 \end{eqnarray}
 The number operators $N_{\alpha}{\equiv}a_{\alpha}^{\dagger}a_{\alpha}$ and
 $N_{\beta}{\equiv}a_{\beta}^{\dagger}a_{\beta}$ are easily 
 constructed from these.
 
 We would like to consider the mixing of one meson states which, for arbitrary meson
 flavor $\alpha$, are given by 
 \begin{equation}
 |{\alpha}{\ra} = a_{\alpha}^{\dagger}|0{\ra} = \cos{\theta}|1{\ra}+\frac{{\gamma}_+\sin{\theta}}{2}|2{\ra}.
 \end{equation} 
 This gives a normalization factor of 
 \begin{eqnarray}
 {\la}{\alpha}|{\alpha}{\ra} &=& \cos^2{\theta}+\frac{{\gamma}_+^2\sin^2{\theta}}{4} \nonumber\\
 				 &=& 1+C_b,
 \end{eqnarray}
 where ${\gamma}_+^2 = 4 + {\gamma}_-^2$ was used and $C_b$ is the boson condensation density given in eq.(12).
 The significance of the normalization will be commented upon later. 
 From the definitions of the number operator and the meson state it is easy to see that 
 \begin{eqnarray}
 {\la}1|N_{\alpha}|1{\ra} &=& \cos^2{\theta}+\frac{{\gamma}_-^2\sin^2{\theta}}{4}
 \nonumber\\
 {\la}1|N_{\alpha}|2{\ra} &=& \frac{{\gamma}_+\cos{\theta}\sin{\theta}}{2}
 \nonumber\\
 {\la}2|N_{\alpha}|2{\ra} &=& \frac{({\gamma}_-^2+{\gamma}_+^2)\sin^2{\theta}}{4}.
 \nonumber\\
 \end{eqnarray}
 From these relations we find
 \bege
 {\la}{\alpha}|N_{\alpha}|{\alpha}{\ra}_N = \cos^2{\theta}+\frac{({\gamma}_+^2+{\gamma}_-^2)\sin^2{\theta}}{4}
 					      = (1+C_b) + C_b, 
 \ende
 where 
 $
  {\la}{\alpha}|N_{\alpha}|{\alpha}{\ra}_N = \frac{{\la}{\alpha}|N_{\alpha}|{\alpha}{\ra}} 
  {{\la}{\alpha}|{\alpha}{\ra}}.
 $
 Similarly, we obtain
 \bege
 {\la}{\alpha}|N_{\beta}|{\alpha}{\ra}_N = \frac{{\gamma}_-^2\sin^2{\theta}}{4}
 					     = C_b.
 \ende
 In order to find formulas for the oscillation of flavors in time we use the time evolution 
 operator given by $U(t) = \exp{(-iH_{1,2}t)}$, where $H_{1,2}|1{\ra}=E_1|1{\ra}$, etc.
 The calculation yields
 \bege
 {\la}{\alpha}(t)|N_{\alpha}|{\alpha}(t){\ra}_N = {\la}{\alpha}|N_{\alpha}|{\alpha}{\ra}_N
 	-\frac{{\gamma}_+^2\cos^2{\theta}\sin^2{\theta}}{1+C_b} \sin^2{\frac{{\Delta}Et}{2}}
 \ende 
 and
 \bege
 {\la}{\alpha}(t)|N_{\beta}|{\alpha}(t){\ra}_N = {\la}{\alpha}|N_{\beta}|{\alpha}{\ra}_N
 	+\frac{{\gamma}_+^2\cos^2{\theta}\sin^2{\theta}}{1+C_b} \sin^2{\frac{{\Delta}Et}{2}}.
 \ende
 We observe that the sum of the number of both species is constant in time, as expected.
 This suggests the interpretation that the oscillation phenomena results from particle flavors
 interacting with the nontrivial vacuum condensation.

 Unlike fermions, multiple bosons can occupy a single quantum state. Thus, we would like
 to see how particle flavors mix in an identically prepared state of n scalar or pseudoscalar
  bosons of flavor $\alpha$ 
 defined by $|n,{\alpha}{\ra}=\frac{(a^{\dagger}_{\alpha})^n}{\sqrt{n!}}|0{\ra}_{1,2}$.
 The calculation is a straightforward generalization of the above methods and the results 
 are
 \bege
 {\la}n,{\alpha}|N_{\alpha}|n,{\alpha}{\ra}_N  
 	       = n(\cos^2{\theta}+\frac{{\gamma}_+^2\sin^2{\theta}}{4})+\frac{{\gamma}_-^2\sin^2{\theta}}{4}
	       = n(1+C_b) + C_b
 \ende
 and 
 \bege
 {\la}n,{\alpha}|N_{\beta}|n,{\alpha}{\ra}_N = \frac{{\gamma}_-^2\sin^2{\theta}}{4} = C_b.
 \ende
 Here the normalization of states is given by
 \bege
 {\la}n,{\alpha}|n,{\alpha}{\ra} = (\cos^2{\theta}+\frac{{\gamma}_+^2\sin^2{\theta}}{4})^n = (1+C_b)^n.
 \ende
 The fact that the states in the ${\alpha},{\beta}$ basis are not already normalized 
 follows from the non-trivial condensation density and the unitary inequivalence of the Fock bases.
 This is observed in Fig.(2), where the total number of particles in a "one" particle state
 is seen to be greater than one. In general, the normalization factor grows exponentially with $n$.
 The preceding equations written in terms of $C_b$ provide a clear and very interesting 
 physical interpretation of the mixing. In eq.(22) the term $C_b$ is simply 
 the static vacuum condensation, whereas the term $n(1+C_b)$ represents a "renormalized"
 number of particles. Each of the $n$ bosons obtains a particle number slightly larger
 than one through its non-perturbative attraction of vacuum condensate. However, this attraction
 of vacuum condensate leaves no holes in the pervasive vacuum condensate, as we still have the 
 static $C_b$ contribution. In a sense, $1+C_b$ just
 redefines what we mean by $1$ particle. This is further verified in eq.(24) where we have the 
 normalization equal to $n$ factors of $1+C_b$, which can be looked at as abstract
 particle number "volume" in Fock space.  
 These results are somewhat different from the naive 
 expectation that putting $n$ bosons in the non-trivial vacuum will yield simply 
 a boson particle number of $n+C_b$. The above results are to be contrasted with the case for fermions {\cite{6}}
 where the authors (eq.(4.13-4.17)) find, after translating into our notation, 
 \bege
 {\la}{\alpha}|{\alpha}{\ra} = 1 - C_f,{\quad} 
 {\la}{\alpha}|N_{\alpha}|{\alpha}{\ra}_N = 1 = (1-C_f) + C_f,{\quad \rm{and}\;\;}
 {\la}{\alpha}|N_{\beta}|{\alpha}{\ra}_N  = C_f,
 \ende
 where $C_f$ is the fermionic condensation density and is of the same form as $C_b$. We see that 
 the particle number in a one particle state is just one. There is no pervasive vacuum condensate nor
 any attracted vacuum condensate, as expected from the exclusion principle. The ${\alpha}$ fermion
 excludes any ${\alpha}$ vacuum condensate, while the ${\beta}$ contribution is entirely condensate.
 The exclusion of condensate can also be seen in the normalization. Time evolution introduces 
 oscillations in both ${\alpha}$ and ${\beta}$ proportional to $n$, for both the fermion and boson case,
 though for fermions $n=1$.
     
 Note in eq.(22-23) that one species is linearly dependent on $n$ while the other is $n$-independent.
 This is very interesting, since it implies that the ratio of the $\alpha$ species 
 to the $\beta$ species grows linearly with $n$. Thus, states with more 
 identically prepared mesons have less mixing "per capita". This is subject to experimental
 test. The relationship does not hold true when the states are allowed to evolve in time : 
 \begin{eqnarray}
 {\la}n(t),{\alpha}|N_{\alpha}|n(t),{\alpha}{\ra}_N &=& 
 {\la}n,{\alpha}|N_{\alpha}|n,{\alpha}{\ra}_N
 -\frac {n{\gamma}_+^2\cos^2{\theta}\sin^2{\theta}}
 { 1+C_b }
   \sin^2{\frac{{\Delta}Et}{2}} \nonumber\\  
 {\la}n(t),{\alpha}|N_{\beta}|n(t),{\alpha}{\ra}_N &=&
 {\la}n,{\alpha}|N_{\beta}|n,{\alpha}{\ra}_N
 + \frac {n{\gamma}_+^2\cos^2{\theta}\sin^2{\theta}}{ 1+C_b }
    \sin^2{\frac{{\Delta}Et}{2}}.
 \end{eqnarray}
 In the static case, we noted that the mixing is related to the vacuum condensation. 
   Dynamically,
  the mixed state further interacts with the vacuum to produce time dependent effects 
  which depend on the number of interacting particles in the mixed state.

 We may also consider the mixing of meson coherent states defined by
 \bege
 |{\cal{C}},{\alpha}{\ra}\;{\equiv}\;{\cal{N}}e^{{\cal{C}}a^{\dagger}_{\alpha}}|0{\ra}_{1,2},
 \ende 
 where ${\cal{C}}$ is a complex number and the normalization is
  ${\cal{N}}=\exp{{\Bigl{[}}\frac{-|{\cal{C}}|^2}{2}(1+C_b){\Bigl{]}}}$.
 Defining $c=\cos{\theta}$ and $s=\frac{{\gamma}_+}{2}\sin{\theta}$, and using the binomial 
 theorem to expand $(a_{\alpha}^{\dagger})^n$ in terms of $a_1$ and $a_2$, we find a useful expression for 
 the coherent state:
 \bege
 |{\cal{C}},{\alpha}{\ra}={\cal{N}}{\sum}_{n=0}^{\infty}{\sum}_{j=0}^n\frac{{{\cal{C}}}^nc^{n-j}s^j}{\sqrt{(n-j)!j!}}|n-j,j{\ra},
 \ende
 where $|n-j,j{\ra}$ represents the state of $n-j$ base eigenstate one particles and $j$ base
  eigenstate two particles.
 Using this, we quickly obtain the following intermediate results : 
 ${\la}{\cal{C}},{\alpha}|N_1|{\cal{C}},{\alpha}{\ra}=|{\cal{C}}|^2c^2$, 
 ${\la}{\cal{C}},{\alpha}|N_2|{\cal{C}},{\alpha}{\ra}=|{\cal{C}}|^2s^2$,
 ${\la}{\cal{C}},{\alpha}|a_{-2}a_{-2}^{\dagger}|{\cal{C}},{\alpha}{\ra} = 1$, 
 ${\la}{\cal{C}},{\alpha}|a_1^{\dagger}a_2|{\cal{C}},{\alpha}{\ra} 
       =  {\la}{\cal{C}},{\alpha}|a_2^{\dagger}a_1|{\cal{C}},{\alpha}{\ra} = cs|{\cal{C}}|^2$.
   With these one can show that 
 \begin{eqnarray}
   {\la}{\cal{C}},{\alpha}|N_{\alpha}|{\cal{C}},{\alpha}{\ra} &=&
    (c^2+s^2)^2|{\cal{C}}|^2 + \frac{{\gamma}_-^2}{{\gamma}_+^2}s^2
   \nonumber\\
   &=& (\cos^2{\theta} + \frac {{\gamma}_+^2} {4} \sin^2{\theta})^2
   |{\cal{C}}|^2 + \frac{{\gamma}_-^2}{4}\sin^2{\theta} = (1+C_b)^2|{\cal{C}}|^2 + C_b
 \end{eqnarray}
 and 
 \begin{eqnarray}
   {\la}{\cal{C}},{\alpha}|N_{\beta}|{\cal{C}},{\alpha}{\ra} &=& \frac{{\gamma}_-^2}{{\gamma}_+^2}s^2
    \nonumber\\
    &=&  \frac{{\gamma}_-^2}{4}\sin^2{\theta} = C_b.
 \end{eqnarray}
 The time dependent relations are derived in a straightforward way and are
   \begin{eqnarray}
   {\la}{\cal{C}}(t),{\alpha}|N_{\alpha}|{\cal{C}}(t),{\alpha}{\ra} 
      &=& (\cos^2{\theta} + \frac {{\gamma}_+^2} {4} \sin^2{\theta})^2
   	  |{\cal{C}}|^2 + \frac{{\gamma}_-^2}{4}\sin^2{\theta}
	  - |{\cal{C}}|^2 \cos^2{\theta}\sin^2{\theta}{\gamma}_+^2 \sin^2{\frac{{\Delta}Et}{2}} \nonumber\\
      &=& (1+C_b)^2|{\cal{C}}|^2 + C_b
          - |{\cal{C}}|^2 \cos^2{\theta}\sin^2{\theta}{\gamma}_+^2 \sin^2{\frac{{\Delta}Et}{2}}
   \nonumber\\
    {\la}{\cal{C}}(t),{\alpha}|N_{\beta}|{\cal{C}}(t),{\alpha}{\ra}
      &=& 
          \frac{{\gamma}_-^2}{4}\sin^2{\theta}
	  + |{\cal{C}}|^2 \cos^2{\theta}\sin^2{\theta}{\gamma}_+^2 \sin^2{\frac{{\Delta}Et}{2}} \nonumber\\
      &=& C_b + |{\cal{C}}|^2 \cos^2{\theta}\sin^2{\theta}{\gamma}_+^2 \sin^2{\frac{{\Delta}Et}{2}}.
   \end{eqnarray}

   Now we show how each of the ladder operators in eq.(14) can be obtained by two similarity
   transformations. First the base eigenstates are rotated together. Then, through a Bogoliubov transformation{\cite{9}},
   the particles are mixed with the antiparticles moving backward in time. We seek operators $R$ and $B$ such that
   \begin{eqnarray}
   a_{\alpha} &=& B_1^{-1}R^{-1}a_1RB_1
   \nonumber\\
   a_{\beta} &=& B_2^{-1}R^{-1}a_2RB_2
   \end{eqnarray}
   We obtain
   \bege
   R=\exp{\Bigl{[}{\theta}{\sum}_{k}({a_{k,1}^{\dagger}a_{k,2}-a_{k,2}^{\dagger}a_{k,1})}\Bigl{]}}
   \ende
   for both mass eigenstate rotations. For the Bogoliubov transformations we need two different operators:
   \begin{eqnarray}
   B_1 &=& \exp{\Bigl{[}{\phi}{\sum}_{k}(a_{k,2}^{\dagger}a_{-k,2}^{\dagger}-a_{k,2}a_{-k,2})\Bigl{]}}
   \nonumber\\
   B_2 &=& \exp{\Bigl{[}-{\phi}{\sum}_{k}(a_{k,1}^{\dagger}a_{-k,1}^{\dagger}-a_{k,1}a_{-k,1})\Bigl{]}},
   \end{eqnarray}
   where $\cosh{\phi}{\equiv}\frac{{\gamma}_+}{2}$ and $\sinh{\phi}{\equiv}\frac{{\gamma}_-}{2}$. 
   
   One should note that the non-trivial mixing phenomena are possible only if both the mixing angle 
   ${\theta}$ is nonzero and the mass difference between the two mass eigenstates does not vanish. As 
   shown in eq.(12), the condensation density of the physical vacuum is nonzero only if 
   these two conditions (${\theta}{\not=}0$ and ${\gamma}_-{\not=}0$) are satisfied. The operators $R$ and
   $B$ given by eqs.(34) and (35) are associated with these two conditions, ${\theta}{\not=}0$ and
   ${\gamma}_-{\not=}0$, respectively. These conditions are required in order for the two operators to be 
   different from the identity operator. Unless both operators are nontrivial (i.e. different from 
   the identity operator), one cannot expect the physically observable mixing phenomena.  

  {\center{\bf {IV. Application to Real Meson States}}}

  To illustrate the results of the previous section we examine the $\eta$ ${\eta}'$ system. 
  The masses are taken to be $549$MeV and $958$MeV, respectively, and of course in the 
  particle rest frame the energies in the above expressions reduce to the masses. 
  The phenomenologically allowed mixing angle (${\theta}_{SU(3)}$) range of the ${\eta} {\eta}'$
  system is given between $-10^o$ and $-23^o${\cite{8}}, where the mixing angle ${\theta}_{SU(3)}$
  is defined by Eq.(36) of Ref.{\cite{9}}. This angle represents the breaking of the SU(3) symmetry,
  the eigenstates of which are already rotated $-35.26^o$ from $u{\bar u}+d{\bar d}$ and $s{\bar s}$
  to ${\eta} = u{\bar u}+d{\bar d}-2s{\bar s}$ and ${\eta}' =
   u{\bar u}+d{\bar d}+s{\bar s}$. Thus,
  our mixing angle is defined by ${\theta}={\theta}_{SU(3)}-35.26^o$.  
  Recent analysis of the $\eta$ ${\eta}'$ mixing angle
  using a constituent quark model based on the Fock states quantized on the light-front can be
  found in Ref.{\cite{10}} and the references therein. The optimal value found for ${\theta}_{SU(3)}$ was
  ${\sim}-19^o$, and thus ${\theta}=-54^o$ was used in generating Fig.1 and Fig.3.
  The $\eta$ ${\eta}'$ system is interesting because it is nearly maximally mixed. In Fig.2
  we see that at $|{\theta}|=45^o$ the time averaged occupation numbers for both particles
  are equal, and are nearly equal in the range of possible $\theta$ values. Fig.1 shows how the 
  flavor oscillations occur on a very short time scale, even compared with the lifetimes of 
  $\eta$ and ${\eta}'$, which are $7{\times}10^{-19}$s and $3{\times}10^{-21}$s, respectively.
  Fig.3 gives the ratio of the quantities plotted in Fig.1.
  
  The same formulation has been applied to the mixing of the $K_L$ $K_S$ system, although
  the CP violation appears to be too minimal to lead to any appreciable meson mixing observables,
  unlike the case
  of the ${\eta}$ and ${\eta}'$ system. However, this issue deserves further investigation.
  
  {\center{\bf{V. Conclusions and Discussions}}}
  
  The non-trivial scalar and pseudoscalar meson mixing effects may be understood by the condensation of 
  corresponding flavor 
  states in the vacuum as presented in this work. Central to this analysis is the interplay 
  between the base (unmixed) Fock space and the physical Fock space. Their nontrivial relationship
  (unitary inequivalence of the vacuum states) gives rise to the mixing and oscillation phenomena.
   While the similar quantum field theoretic formulation
  was presented for the fermion mixing {\cite{6}}, our analysis intrinsically differs from the fermion case 
  because of the fundamental difference in statistics. As a consequence, we found that the unitary 
  inequivalence of the base flavor states and the physical mass eigenstates holds even in the finite 
  volume regime, in contrast to the case of fermion mixing where the unitary inequivalence holds only 
  in the infinite volume limit{\cite{6}}.
  An interesting physical interpretation of the results is that an $n$ boson state can be thought 
  of as a sum of the static vacuum condensate, a "renormalized" number of bosons $n(1+C_b)$, and 
  time evolution effects.   
  We also noted that, for both the 
  boson and fermion cases, the non-trivial observable mixing phenomena cannot occur unless there is both
  a nonzero mixing angle and also a nonzero mass (energy) difference between the two physically 
  measurable mixed states.
  	
	 As a physical application, we used our formulation to analyze the 
  ${\eta}$ ${\eta}'$ system and found that the measured mixing angle and mass difference between 
  ${\eta}$ and ${\eta}'$ can be related to the non-trivial flavor condensation in the vacuum.
  However, more fundamental questions such as the translation of the condensation in hadronic 
  degrees of freedom to those in quark and gluon degrees of freedom remains unanswered. The answer to
  this question depends on the dynamics responsible for the confinement of quark and gluon degrees of
  freedom and perhaps has to rely on lattice QCD and/or some phenomenological model that accomodates 
  strongly interacting QCD. Further investigation along this line is underway. Also, it would be interesting 
  to look at the mixing transformations between gauge vector bosons governed by the Weinberg angle in the
  electroweak theory as well as vector mesons such as the $\rho$ and $\omega$.
   While the statistics are the same as the scalar and pseudoscalar bosons considered here, there will
  be additional spin dependent interactions which complicate the analysis.  
 
    \sizeplot{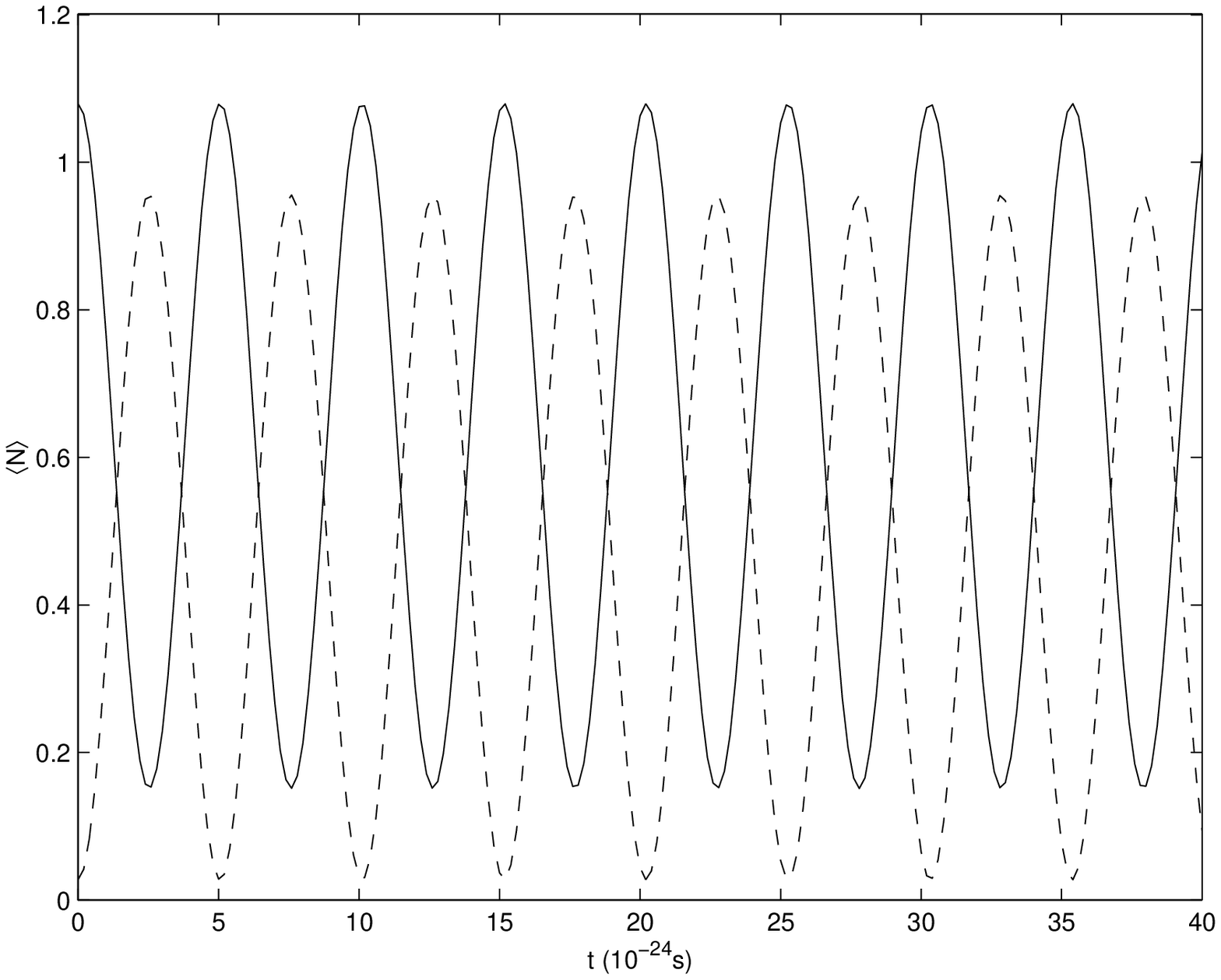}{}{4.0in}{3.5in}{htb}{The expectation value of the number operator for $\alpha=\eta$ and 
	$\beta=\eta'$ in a $n_{\eta}=1$ state. The solid and dashed curves correspond 
	to ${\la}n_{\eta}(t)|N_{\eta}|n_{\eta}(t){\ra}_N$ and 
	 ${\la}n_{\eta}(t)|N_{\eta'}|n_{\eta}(t){\ra}_N$, respectively, as given by 
	   Eq.25. The mixing angle is taken to be $\theta=-54^o$ }
    \sizeplot{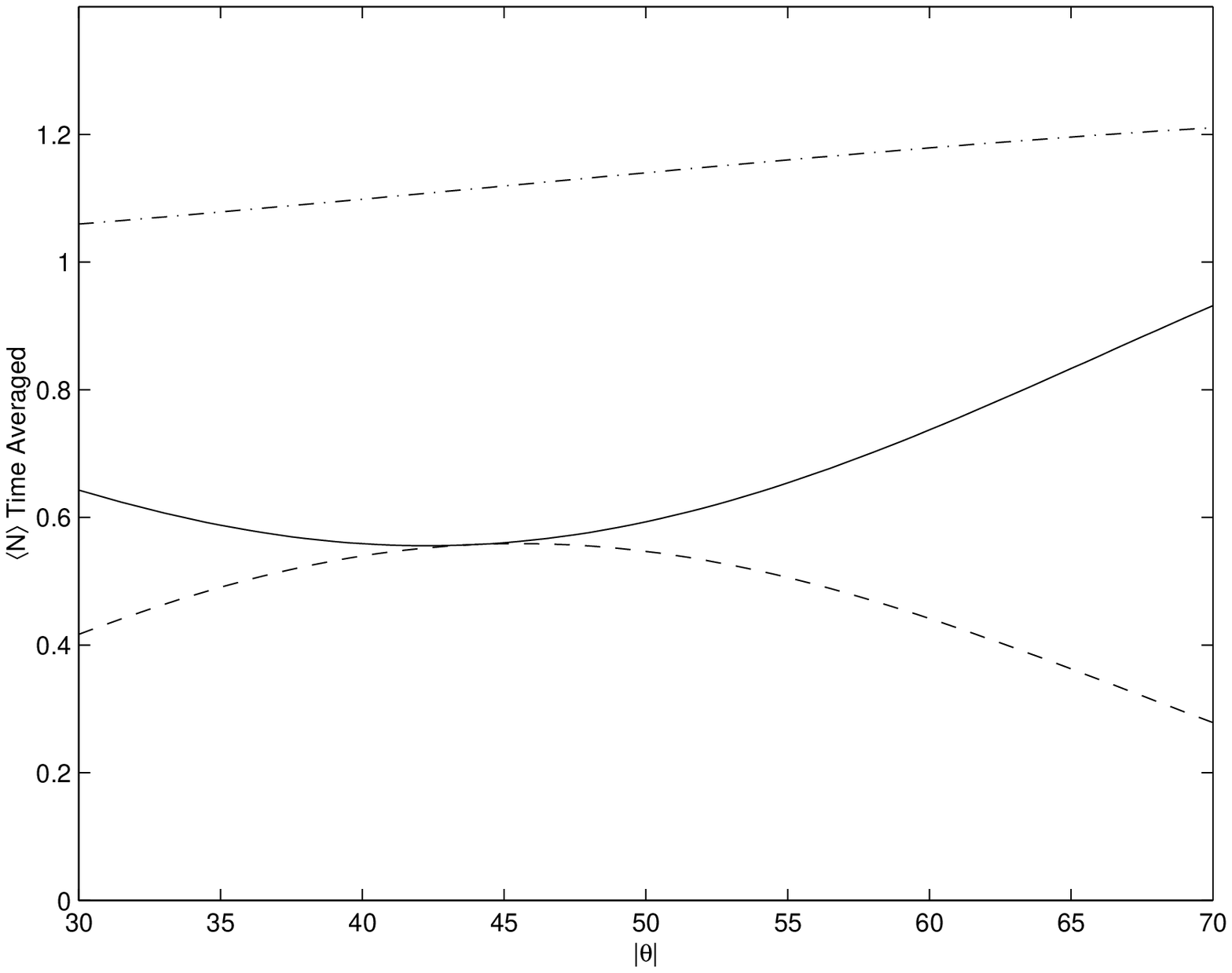}{}{4.0in}{3.5in}{htb}{The time averaged occupation number expectation values for 
 	for the $n_{\eta}=1$ state plotted versus $|{\theta}|$, the mixing angle.
 	The solid and dashed lines represent the time averaged values 
 	of ${\langle}N_{\eta}{\rangle}$ and ${\langle}N_{{\eta}'}{\rangle}$, respectively. The dash-dotted 
 	line is the sum of the two.}
    \sizeplot{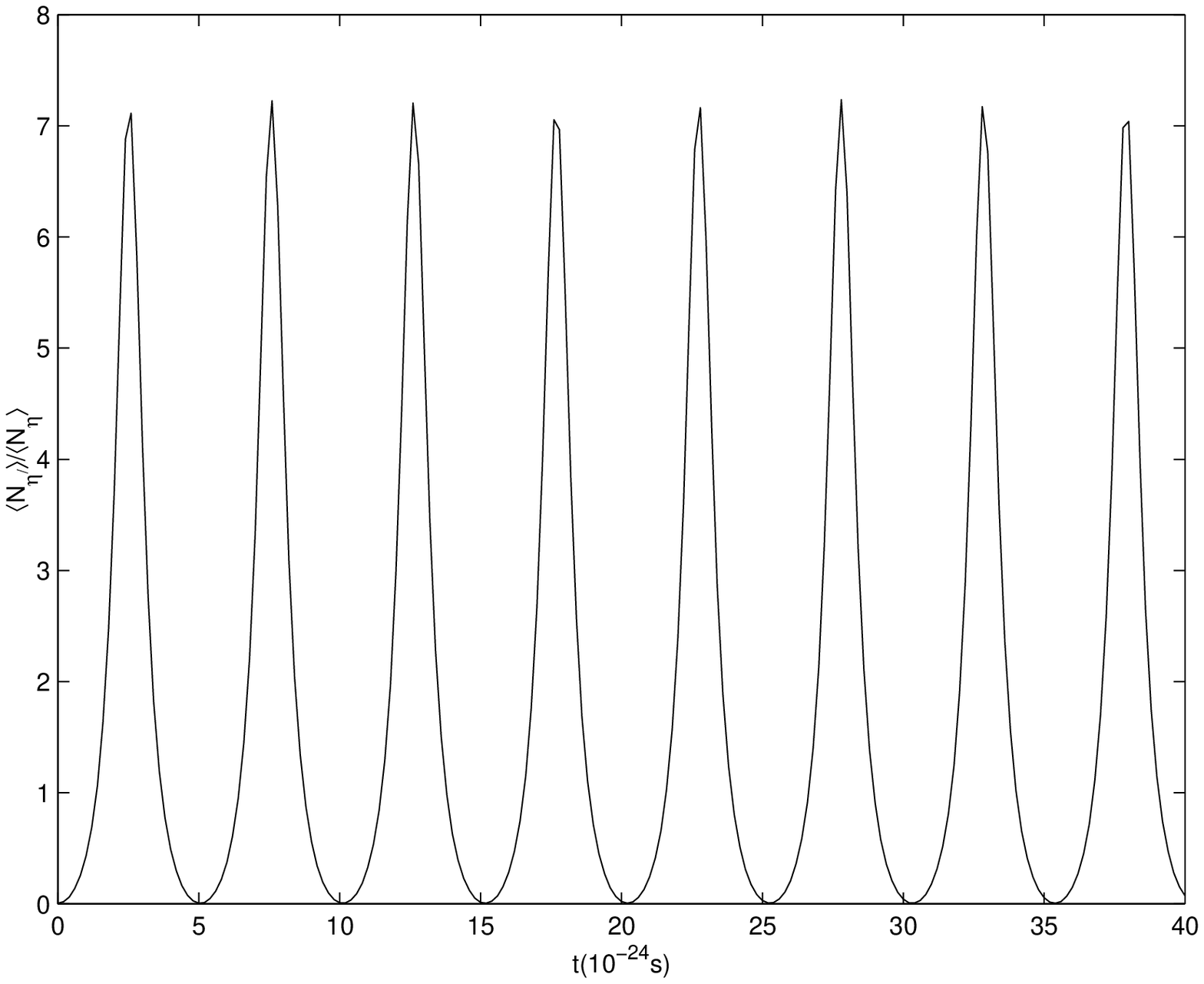}{}{4.0in}{3.5in}{htb}{The ratio of the expectation values of the number operators
	 for $\eta$ and $\eta'$, as given by Eq.25, for an arbitrary $n_{\eta}$ state.
	 The value $n_{\eta}$ is unimportant since any value will yield an almost 
	 identical curve, with the $n_{\eta}=1$ case only being shifted down slightly,
	 reflecting the relative abundance of vacuum condensation.}

\end{document}